\documentclass[structabstract]{aa}  
\usepackage{graphicx}
\usepackage{longtable}
\usepackage[authoryear]{natbib}
\bibpunct{(}{)}{;}{a}{}{,}
\usepackage{txfonts}
\begin{document}
\title{Another cluster of red supergiants close to RSGC1}
\author{
I.~Negueruela\inst{1}
\and C.~Gonz\'alez-Fern\'andez\inst{1}
\and A.~Marco\inst{1}
\and J. S. Clark\inst{2}
\and S.~Mart\'{\i}nez-N\'u\~{n}ez\inst{1}}

\institute{
Departamento de F\'{i}sica, Ingenier\'{i}a de Sistemas y
  Teor\'{i}a de la Se\~{n}al, Universidad de Alicante, Apdo. 99, E03080
  Alicante, Spain\\
\email{ignacio.negueruela@ua.es}
\and
Department of Physics and Astronomy, The Open 
University, Walton Hall, Milton Keynes, MK7 6AA, UK}

\abstract{Recent studies have revealed massive star
  clusters in a region of the Milky Way close to the tip of the Long
  Bar. These clusters are heavily obscured and are characterised by a
  population of red supergiants.}
{We analyse a previously unreported concentration of bright red stars
  $\sim16\arcmin$ away from the cluster RSGC1}
{We utilised near IR photometry  
to identify candidate red supergiants and then $K$-band spectroscopy of a
sample to characterise their properties.}
{We find a compact clump of eight red supergiants and five other
  candidates at some distance, one of which is spectroscopically
  confirmed as a red supergiant. These objects must form an open cluster, which we name Alicante~8. Because of the high reddening and strong field
  contamination, the cluster sequence is not clearly seen in 2MASS or UKIDSS
  near-IR photometry. From the analysis of the red supergiants, we infer an extinction
  $A_{K_{{\rm S}}}=1.9$ and an age close to 20~Myr.}{Though this cluster is
  smaller than the three known previously, its properties still suggest a mass in excess of
  $10\,000\:M_{\sun}$. Its discovery corroborates the hypothesis that
  star formation 
in this region has happened on a wide scale between $\sim10$ and
$\sim20$~Myr ago.} 
 

\maketitle

\section{Introduction}

\defcitealias{davies07}{D07}

Over the past few years, the census of massive ($M_{{\rm
    cl}}\ga10^4M_{\sun})$ clusters in the Milky Way has steadily
increased, with the discovery of three such clusters near the Galactic
centre  \citep{krabbe,nagata,cotera,figer99} and the realisation that
\object{Westerlund~1} has a mass of the
 order of $10^5M_{\odot}$  \citep{clark05}. Similar clusters are known
 in many other galaxies and are typical of starburst environments,
 where they appear in  extended complexes
 \citep[e.g,][]{nate05}. Targeted searches revealed three more massive clusters in a small region of the Galactic plane,
 between $l=24\degr$ and $l=29\degr$ \citep{figer06,davies07,
   clark09}. The Long Galactic Bar is believed to end in this region
 \citep{cabrera08}, touching what has been called the base
 of the Scutum-Crux arm (\citealt{davies07}, from now on \citetalias{davies07}), which may also be considered
 a part of the Molecular Ring \citep[e.g.,][]{rathborne09}.

These three highly-reddened clusters are dominated by large
populations of red supergiants (RSGs), which appear as very bright
infrared sources, while their unevolved populations have not been yet
characterised. RSGC1 is the most heavily obscured, with an estimated
$\tau=12\pm2\:$Myr and $M_{{\rm initial}}=3\pm1\times10^4M_{\sun}$
\citep{davies08}. RSGC2 = Stephenson 2 is the less obscured and
apparently most massive of the three, with $\tau=17\pm3$~Myr and $M_{{\rm
    initial}}=4\pm1\times10^4M_{\odot}$ \citepalias{davies07}.  Finally,
RSGC3 lies at some distance from the other two and has an estimated
$\tau=16$--20\,Myr and an inferred $M_{{\rm initial}}=
 2$--$4\times10^4M_{\sun}$  \citep{clark09,alexander09}. Collectively,
 the three clusters are believed to host $>50$ true RSGs (i.e.,
 $M_{{\rm ZAMS}}\ga12\:M_{\odot})$, the kind 
of objects thought to be the progenitors of Type IIn supernovae
\citep{smartt}.
 
\begin{figure*}
\label{fig:colour}
\resizebox{\textwidth}{!}{
\includegraphics[angle=90]{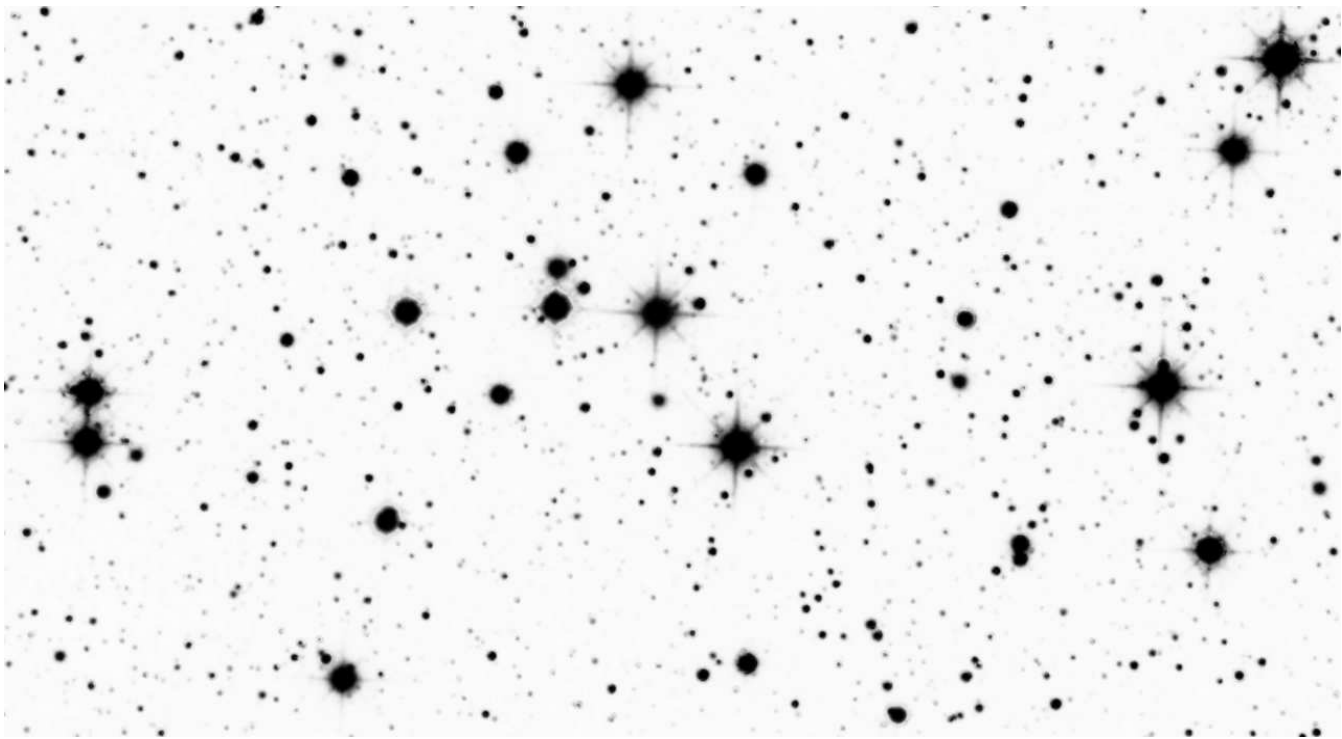}}
\caption{Near-IR $JHK$-band colour composite  of the field around
  Alicante~8, constructed from  
UKIDSS data with artifacts due to saturation artificially
removed and colour enhancement. Note the lack of a clearly defined stellar  
overdensity of unevolved cluster members with respect to the
field. The image covers approximately $5\farcm5\times4\arcmin$.}
\end{figure*}

In this paper, we report the discovery of one more cluster of red supergiants in the immediate vicinity of RSGC1, which we
designate as Alicante~8 = RSGC4\fnmsep\footnote{Though designating this cluster RSGC4 may seem the most natural step, this choice raises the question of when a cluster should be considered a cluster of red supergiants, i.e., how many red supergiants are needed and how prominent the supergiants have to be with respect to the rest of the cluster. For this reason, we favour the alternative names.}.
Identified visually in 2MASS $K_{{\rm S}}$ images as a concentration
of bright stellar sources near $\ell=24\fd60, b=+0\fd39$ (see
Fig.~\ref{fig:colour}),  
we  utilised near-IR photometry to identify  potential cluster
members, nine of which were subsequently observed
spectroscopically and confirmed to be RSGs. Though this cluster is
perhaps less massive than the other three, it
provides further evidence for the presence of an extended star
formation region in the direction of the end of the Long Bar.

\section{Data acquisition  and reduction}

\begin{figure}
\label{fig:finder}
\resizebox{\columnwidth}{!}{
\includegraphics[angle=-0]{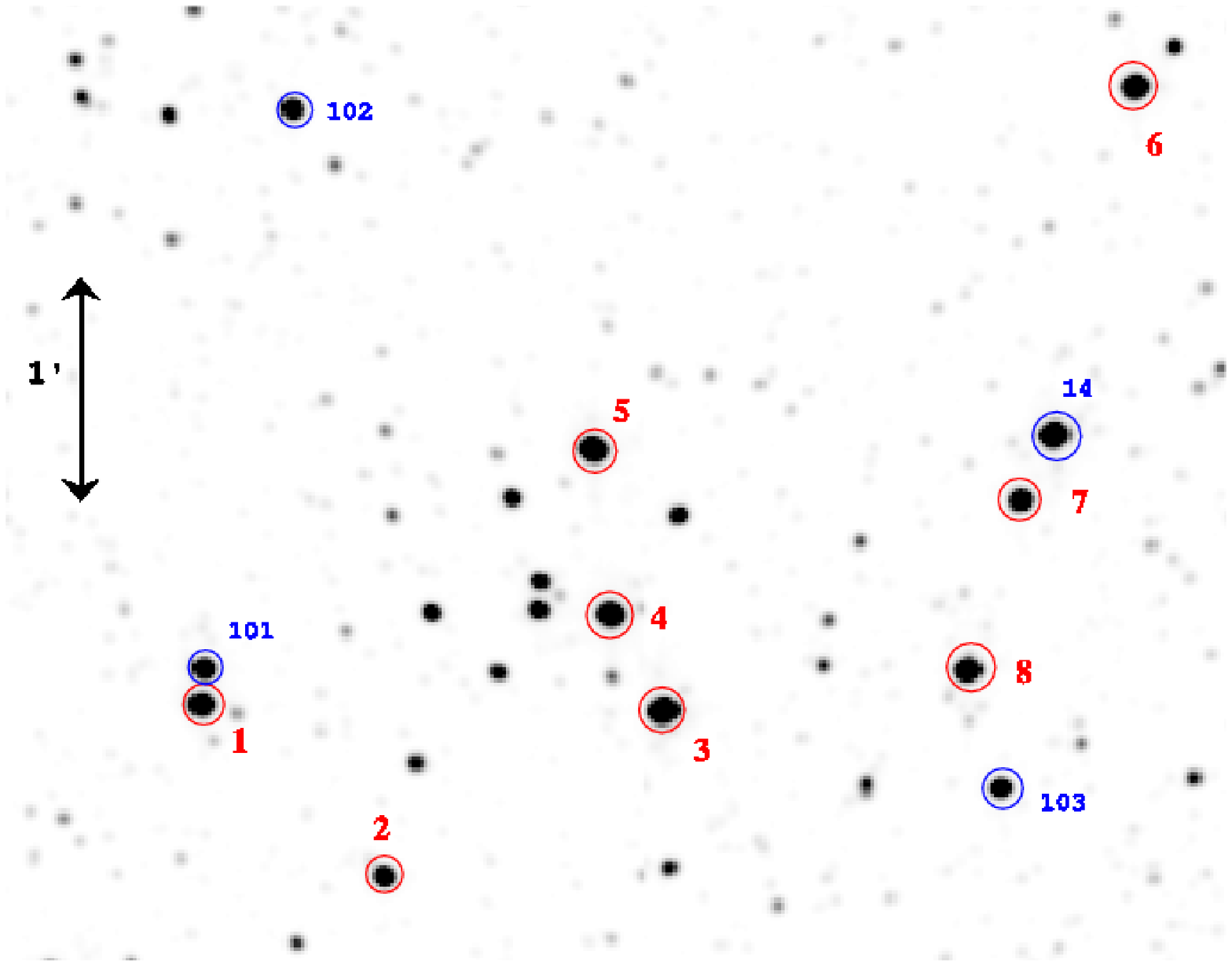}}
\caption{Finding chart for Alicante~8, with the stars listed in Table~1
  indicated. The finder comprises a $K$-band image from 2MASS with a
  $\sim7\arcmin\times5\arcmin$ field encompassing all the confirmed
  members (S1--8, marked by red circles). Other stars discussed in the
text are marked by blue circles.}
\end{figure}

As discussed by \citet{clark09}, it is extremely difficult to determine  
 a physical extent for any of the RSG clusters, since their unevolved
 populations are not readily visible 
 as overdensities with respect to the field
 population in any optical or infrared band. In view of this, we must
 rely on the apparent concentration of 
 bright infrared sources to define a new cluster. In the absence of
 spectral and/or kinematical information, it is difficult to
 distinguish between {\it bona fide} cluster RSGs 
and a diffuse field population \citepalias[cf.][]{davies07}. We are thus
forced to utilise  photometric data to 
construct  a list of candidate cluster members.

\subsection{2MASS data}
\label{sec:2mass}

 We have used 2MASS $JHK_{{\rm S}}$ photometry to identify the RSG
 population.  Based on the spatial concentration of bright red star
 (Fig.~\ref{fig:finder}), 
 we start by taking 2MASS photometry for stars within $r\leq7\arcmin$
 of the position of Star~4 (RA: 18h 34m 51.0s,
 DEC:$-07\degr\:14\arcmin\:00\farcs5$), selecting stars with low
 photometric errors ($\Delta K_{{\rm S}}\leq0.05$). A number of bright
 stars defining the spatial concentration have very high  $(J-K_{{\rm
     S}})\approx3.5$ values and form a well-separated clump
in the $(J-K_{{\rm S}})/K_{{\rm S}}$ diagram (Fig.~3). The
clump, which comprises 11 stars, is
also very well defined in the $(H-K_{{\rm S}})/K_{{\rm S}}$ diagram,
centred around $(H-K_{{\rm S}})=1.2$. We name these stars S1--3 and
S5--12. 

We make use of the reddening-free parameter
$Q=(J-H)-1.8\times(H-K_{{\rm S}})$ \citep[see, e.g.,][]{ns07} to
estimate the nature of 
stars. Using, e.g., the intrinsic colour calibration of
\citet{straizys09}, we see that early-type stars must have $Q\la0.0$,
while the dominant population of bright field stars, red  
clump giants, have $Q\approx 0.4-0.6$. Perhaps because of colour terms
and the structure of their atmospheres, most RSGs do not deredden
correctly when the standard law is assumed, and give values
$Q=0.1-0.4$. Examination of the fields of the three known RSG clusters shows that more than two thirds of the RSGs give low values of $Q$ ($\approx 0.1$ -- 0.3), while the remaining show $Q\approx0.4$, typical of red stars. No dependence with the spectral type is obvious.

Of the 11 stars in the clump, 9 have $Q$ in the typical range for
RSGs. One other object, S8 has its $J$ magnitude marked as unreliable in
2MASS, and has therefore an unreliable $Q$ value. The final star, S12,
has $Q=-0.15$, indicative of an infrared 
excess. In addition to these 11 objects, two other stars with $Q$ in
the interval typical of 
supergiants, S4 and S13, have redder $(J-K_{{\rm S}})$ and
$(H-K_{{\rm S}})$  colours than the rest. We consider the 11 stars in
the clump and these two redder stars as candidate RSGs. Finally, one
star S14, has $Q$ typical of supergiants, but much bluer colours, and
we do not consider it a candidate cluster member, but a candidate
foreground RSG. Stars S1--8 are spatially concentrated and define the
cluster core (Fig.~2). Stars S9--13 are located at greater distances, and not
shown in Fig.~2. The coordinates and magnitudes of all the stars under
discussion are listed in Table~\ref{tab:props}.

\begin{figure}
\label{fig:cmd}
\resizebox{\columnwidth}{!}{\includegraphics[angle=-0,clip]{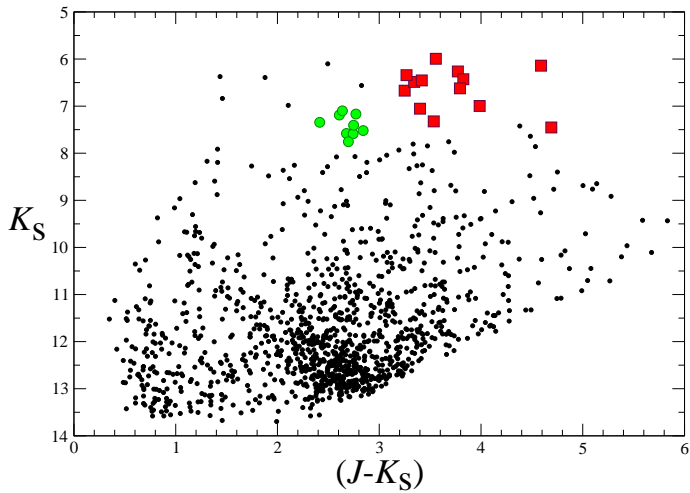}}
\caption{Colour magnitude plot for stars within $7\arcmin$ of
Alicante~8, using 2MASS data. The likely cluster members identified in
Sect.~\ref{sec:2mass} are   
indicated by the red squares, while the group of less luminous objects
discussed in Section~\ref{sec:discu} are plotted as green circles.  
 Note that the two stars with $(J-K_{{\rm S}})\sim4.5$ are S4
 and S11 (see text). The former is spectroscopically confirmed as an RSG, but the
 second is fainter than most members, and requires spectroscopic
 study.} 
\end{figure}

\subsection{Spectroscopy}

A sample of the candidates were subsequently observed with the
Long-slit Intermediate Resolution Infrared Spectrograph
(LIRIS) mounted on the 4.2~m William Herschel Telescope (WHT), at the
Observatorio del Roque de los Muchachos (La Palma, Spain). The
instrument is equipped with a $1024\times1024$ pixel HAWAII detector. Stars 1,
3--6 \& 8 were observed in service mode on the night of June 29,
2009, while stars 2, 7 \& 9 were observed during a run on July 6 \& 7,
2009. The configuration was the same in both cases. We profited from the
excellent seeing to use the $0\farcs65$ slit in combination with the
intermediate-resolution $K$ pseudogrism. This combination covers the
2055--2415~nm range, giving a
minimum $R\sim2500$ at 2055~nm and slightly higher at longer
wavelengths. 

Data reduction was carried out using dedicated software developed by
the LIRIS science group, which is implemented within
IRAF\fnmsep\footnote{IRAF is 
  distributed by the National Optical Astronomy Observatories, which
  are operated by the Association of Universities for Research in
  Astronomy, Inc., under cooperative agreement with the National
  Science Foundation}. We used the A0\,V star HIP 90967
to remove atmospheric features, by means of the
{\sc xtellcor} task \citep{vacca03}.  The
spectra of all the stars are shown in Fig.~4. We also
shown the spectrum of a star which felt by chance inside the slit when
observing S1. We call this star S101 and will discuss it further down.

\begin{figure*}
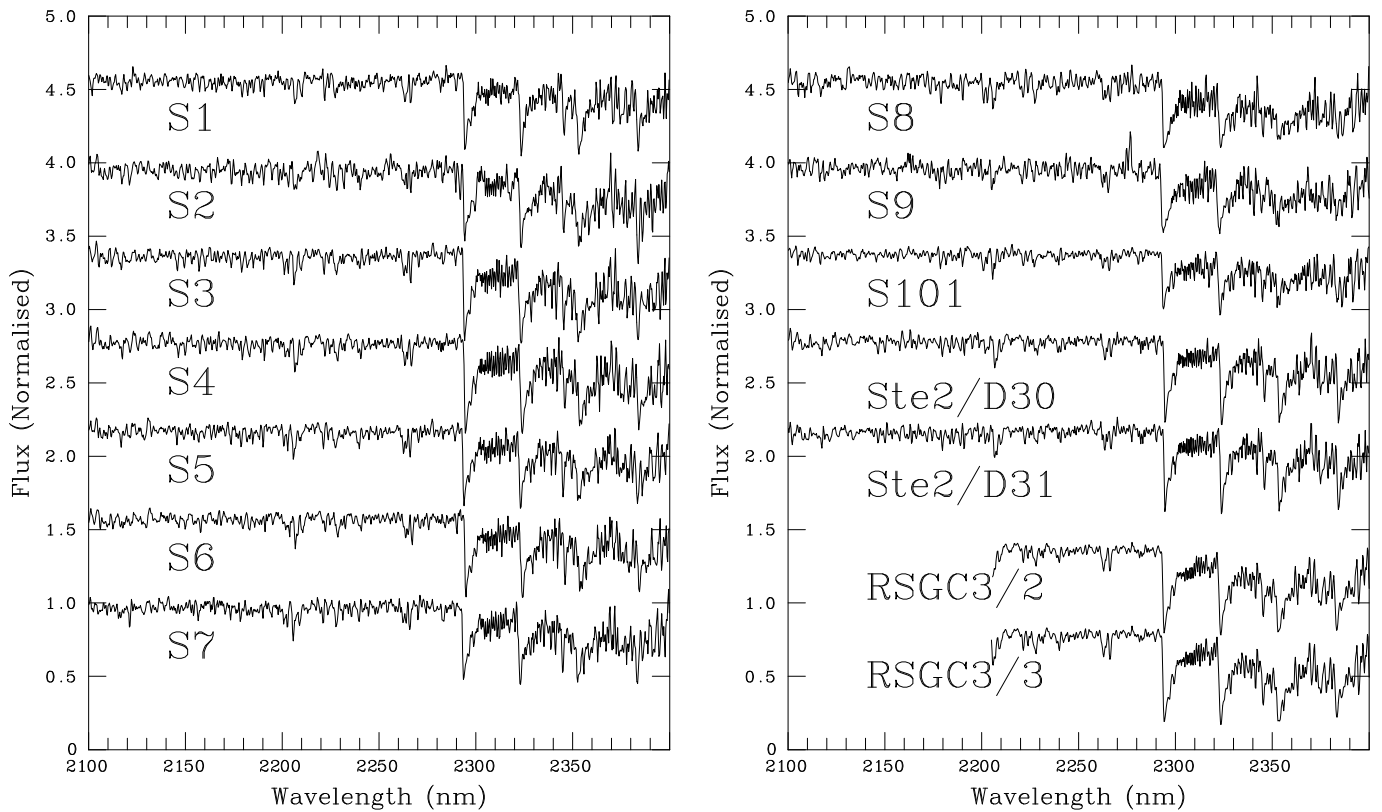
 
\label{fig:spec}\resizebox{\textwidth}{!}{
\includegraphics[width=8cm]{13373fg4a.eps}
\includegraphics[width=8cm]{13373fg4b.eps}}
\caption{{\bf Left: } $K$-band spectra of the eight confirmed members
  in the core of 
  the new cluster, Alicante~8. {\bf Right: } $K$-band spectra of two
  other stars in the field. S9 is a supergiant at some distance from
  the cluster, which may well be a member, in spite of slightly lower
  reddening.  S101 is a foreground bright giant coincident with the
  cluster,  part of a population spread over the whole field. As a
  comparison, we show two RSGs in 
  Stephenson~2 observed with the same setup. We also show two RSGs in
  RSGC3, observed at similar resolution \citep{clark09}.}
\end{figure*}

\subsection{UKIDSS data}

We complete our dataset by utilising UKIDSS $JHK$ photometry
\citep{lawrence07}.  The data 
were taken from the Galactic Plane Survey \citep{lucas08} as provided
by the Data Release 4 plus.

\section{Results}

\subsection{Supergiant members}

Figure~4 shows the spectra of candidates S1--9, together
with that of S101.  All the stars
observed show deep CO bandhead
absorption, characteristic of late type stars.  Following the
methodology of \citetalias{davies07}, it is possible to use the equivalent
width of the CO bandhead feature, EW$_{{\rm CO}}$, to provide an
approximate spectral classification for the stars.

\citetalias{davies07} measure the EW$_{{\rm CO}}$ between 2294--2304~nm. Unfortunately, at the resolution and signal-to-noise of our spectra, the continuum band defined by \citetalias{davies07} does not provide a reliable determination of the continuum. Therefore we choose to select the continuum regions from \citet{carlos08}, with
which this value is obtained over a wider range in wavelength and
therefore less prone to be tainted by spurious effects.
We use the spectra of two confirmed RSGs in Stephenson~2
with magnitudes comparable to our sample (observed with the same
setup) to ensure that our EWs are
measured in the same scale as those of \citetalias{davies07}. The values measured agree within 1\AA\ with those determined by \citetalias{davies07}.

In addition, we profit from the recent publication of the atlas of infrared spectra of \citet{rayner09} to verify the calibration of spectral type against EW$_{{\rm CO}}$ \citepalias{davies07}. Thanks to the atlas, we can use a much higher number of M-type stars than in the original calibration and extend it to later spectral types. We measure EW$_{{\rm CO}}$ by defining the same continuum regions as used for our targets. The results are plotted in Fig.~5.

\begin{figure*}
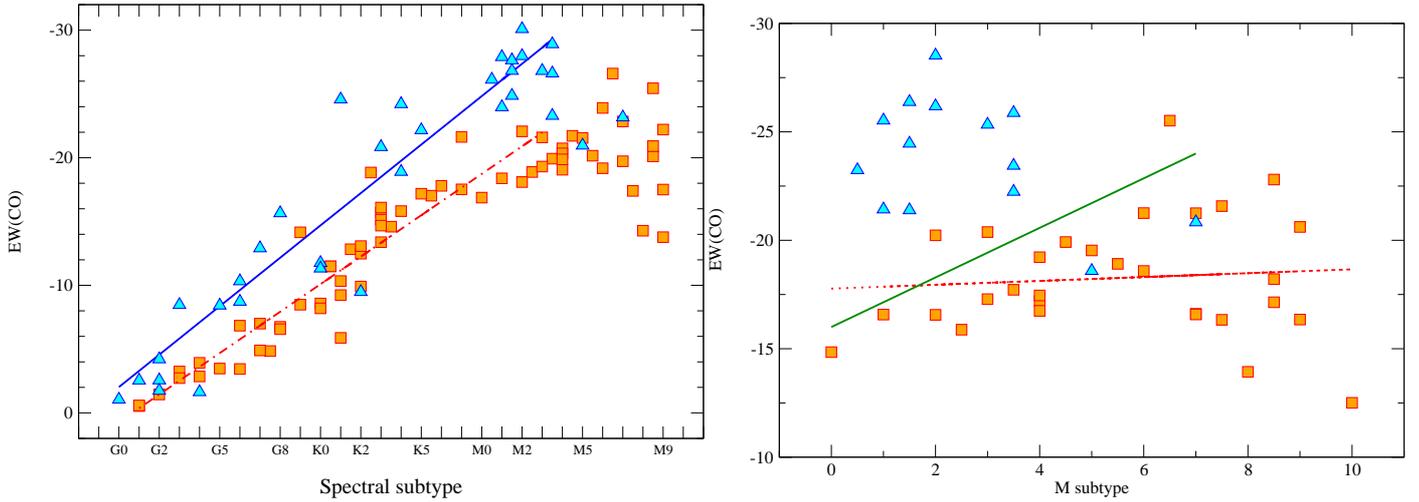

\label{fig:daviesrel}
\resizebox{\textwidth}{!}{\includegraphics[angle=-0,clip]{13373fg5a.eps} \includegraphics[angle=-0,clip]{13373fg5b.eps}}
\caption{{\bf Left panel:} Relationship between spectral type and the equivalent width of the CO bandhead for G--M type stars in the catalogue of \citet{rayner09}. Giants are plotted as squares while supergiants are triangles. The continuous line is a fit to all the supergiants between G0 and M3, with the exception of two spectrum variables mentioned in the text. The dot-dashed line is the fit to all giants between G0 and M3. {\bf Right panel:}
Same for only the M-type stars in the catalogue of \citet{rayner09}, including some Mira-type spectrum variables which were excluded from the left panel. The dotted line represents the best fit to the data for all the M-type giants, while the continuous line is the fit by \citetalias{davies07} to giants in the range G0 to M7.
} 
\end{figure*}

In the plot, we have used all the giants and supergiants with spectral type between G0 and M7, leaving out a few early G objects with no measurable CO bandhead. We have also included giants with spectral type M8--9. For G and K stars, our results reproduce very well those of \citetalias{davies07}. Supergiants and giants appear well separated, with a few exceptions. Some of the exceptions are due to spectral variability. For instance, two of the three supergiants falling close to the position of the giants are known spectral type variables (RW~Cep and AX~Sgr), and have not been used for the fit. The two giants falling along the location of the supergiants have luminosity class II. Most objects with luminosity class II have higher EW$_{{\rm CO}}$ than luminosity class III objects of the same spectral type, but only these two stand out strongly.

For the M stars, the situation is not so clear. At a given spectral type, there is very significant scatter in the values of EW$_{{\rm CO}}$, especially for supergiants, but also for very late giants. Most supergiants have higher EW$_{{\rm CO}}$ than most giants, but there are a few exceptions in both directions. This is, in part, not so surprising, because some AGB stars are as luminous as some supergiants \citep{vanloon05}. Our sample almost completely lacks supergiants later than M4. The apparent lack of correlation between EW$_{{\rm CO}}$ and spectral type for M supergiants is partly due to the position of the M5\,Ib--II star HD~156014, which has a very low luminosity, and is the only supergiant in the range. As \citetalias{davies07} have several supergiants with spectral types $>$M4, we will accept their calibration in this range.

Our data show that the slope of the relationship does not keep constant for giants with spectral type $\geq$M5, as these objects do not show, on average, higher values of EW$_{{\rm CO}}$ than the earlier M giants. This is comforting, as it supports the assumption -- based on the calibration of \citetalias{davies07} -- that any star with EW$_{{\rm CO}}>24$\AA\ is almost certainly a supergiant, and that any star with EW$_{{\rm CO}}\ga22$\AA\ is very likely a supergiant.

Turning back to our targets, S101, which
was not selected as an RSG candidate, shows EW$_{{\rm CO}}=18$\AA, a value  typical of M giants. All the other stars have higher equivalent widths, in the region of supergiants. In particular, stars S3, S4, S6, S8 and S9 have  EW$_{{\rm CO}}>24$\AA, and must be RSGs according to the calibration of \citetalias{davies07}. The other four stars have $21$\AA~$<$~EW$_{{\rm CO}}<24$\AA\ and can be either K supergiants or M giants. Of them, only S2 has a $Q$ value compatible with being a red giant. Based on this, we assume that all the candidates are supergiants, though noting that S2 could be a giant.

As discussed, we use the calibration of \citetalias{davies07} to estimate spectral types
for the confirmed supergiants. The derived types, which must be
considered approximate because of the procedure used (\citetalias{davies07} estimate uncertainties of $\pm2$ subtypes), are listed in
Table~\ref{tab:sgs}. Interestingly, S4, which has the redder colours,
also has the deepest CO bandhead, indicative of a spectral type
M6\,I. Though the spectral types are approximate, the distribution is not very different from the other RSG clusters, with types extending from K4\,I to
M6\,I. We note that there seems to be some tendency for lower mass RSGs to have spectral type K \citep{hme84,levesque}.

Further, we calculate the $Q$ value for all the stars in the atlas of \citet{rayner09}, finding that almost all K and M-type giants have $Q\approx 0.4$--0.6, with the exception of Miras, which have $Q<0$ because of the colour excess caused by their dust envelopes. This also supports a supergiant nature for all our likely members (S2 may still be a red giant, but it falls together with the other members in the photometric diagrams).

The eight candidates in the central concentration are very likely all
RSGs, and thus we take them as cluster members, even in
the absence of kinematic data. Of the halo candidates, we have only
observed S9. This object is slightly less reddened than the confirmed
members. As seen in Fig.~4, its morphology resembles more
closely that of S101 than those of the confirmed RSGs. However, the
measured EW$_{{\rm CO}}=25$\AA\ indicates that this object must definitely
be a supergiant, though we cannot confirm its membership,
as we lack kinematic data.

\begin{table*}
\caption{Summary of RSG candidates and their properties$^{a}$. {\bf Top
    panel: } Spectroscopically confirmed RSGs and photometric
  candidates without spectroscopic observations. {\bf Bottom panel:
  }Other objects whose photometric properties are indicative of
  luminous red stars, but are likely to be foreground to the cluster.
 The last column gives the offset between the star location and the
nearest {\it MSX} source$^{b}$.}
\label{tab:props}
\begin{tabular}{lcccccccccccccc}
\hline
\hline
ID & \multicolumn{2}{c}{Co-ordinates}  & \multicolumn{4}{c}{2MASS} & \multicolumn{3}{c}{GLIMPSE   {\it
  Spitzer}$^{\,c}$} & \multicolumn{4}{c}{{\it MSX}} & Offset \\
   &      R.A.     &    Dec.  &   $J$ & $H$    & $K_{{\rm S}}$ &  $Q$ & 4.5$\mu$m & 5.8$\mu$m  & 8.0$\mu$m  &  A   &   C  &   D   &  E &  \\
\hline
&&&&&&&&\\
S1  & 18 34 58.40 &$-$07 14 24.8 & 	 9.92 &    7.80  &	 6.67  &  0.10 &  $-$ & 5.54 & 5.32 & 5.17 & 4.48 & 4.29 &  $-$ &  $3\farcs2$\\
S2  & 18 34 55.12 &$-$07 15 10.8 & 	 10.86 &    8.45  &	 7.32 & 0.38 & 6.73 & 6.37 & 6.31 & 6.07 & 5.11 & 5.05 &  $-$ &  $2\farcs1$ \\
S3  & 18 34 50.00 &$-$07 14 26.2 & 	 9.55  &    7.19  &	 6.00  &  0.22 &  $-$ & 4.84 & 4.68 & 4.69 & 3.57 & 3.42 &  3.32& $2\farcs8$\\
S4  & 18 34 51.02 &$-$07 14 00.5 & 	 10.73 &    7.68  &	 6.14  & 0.29 &  $-$ & 4.65 & 4.50 & 4.45 & 3.43 &3.35 &3.02 & $0\farcs5$\\
S5  & 18 34 51.33 &$-$07 13 16.3 & 	 10.04 &    7.53  &	 6.26 & 0.23& \multicolumn{3}{c}{none within $10\arcsec$} & 5.00 & 4.01 & 3.87 &  $-$  & $1\farcs5$\\
S6  & 18 34 41.55 &$-$07 11 38.8 & 	 10.26 &    7.73  &	 6.43  & 0.18 & $-$  & 5.30 & 5.19 & 5.14 & 4.49 & $-$ &  $-$ &  $2\farcs7$  \\
S7  & 18 34 43.56 &$-$07 13 29.7 & 	 10.46 &    8.22 &	 7.06  & 0.13 & $-$  & 5.94 & 5.88 &   \multicolumn{5}{c}{none within $10\arcsec$}\\
S8  & 18 34 44.51 &$-$07 14 15.3 & 	 10.42:$^{c}$ &    8.03  &	 6.62  & $-0.15^{d}$:&  $-$ &  4.75 &4.63 & 4.34 &3.56 &3.18  &2.86  & $0\farcs7$\\
S9  & 18 34 45.81 &$-$07 18 36.2 & 	 9.83  &    7.63  &	 6.49  & 0.14&  $-$ &  4.99 &4.83 & 4.96 &4.48 &  $-$ &  $-$ &  $1\farcs7$\\
S10 & 18 34 26.81 &$-$07 15 27.9 & 	 10.98 &    8.32  &	 7.00  & 0.29 &  $-$ &  5.78 &5.61 & 5.22 &4.12 &  $-$ &  $-$ &  $0\farcs9$ \\
S11 & 18 35 00.32 &$-$07 07 37.4 & 	 9.61  &    7.47  &	 6.34  & 0.11 &  $-$ &  5.35 &5.12 & 4.73 &3.69 &3.59 &2.77 & $0\farcs3$\\
S12 & 18 35 16.88 &$-$07 13 26.9 & 	 9.87  &    7.73  &	 6.45  & $-0.15$ &  $-$ &  4.66 &4.45 & 4.54 &3.48 &3.31 &3.39 &$0\farcs5$\\
S13 & 18 35 10.89 &$-$07 15 17.8 & 	 12.14 &    9.09  &	 7.45  & 0.11 & \multicolumn{3}{c}{none within $7\arcsec$} & 4.16 &3.08 &2.81 &2.26 &$1\farcs0$  \\
\hline 
&&&&&&&&\\
S14  & 18 34 42.94 &$-$07 13 12.2 &   	 8.60  &    6.87 &   6.10 &  0.35 &  $-$ &  5.53 &5.50 &  5.23 & 4.69 & 5.12& 1.21& $>6\arcsec$\\
S101 & 18 34 58.36 &$-$07 14 15.1 &   	 9.80  &    7.96 &   7.19 & 0.45&  $-$ &  6.61& 6.57&  5.17& 4.48& 4.20&  $-$ & $>6\arcsec$ \\
S102 & 18 34 56.77 &$-$07 11 44.9 &   	 10.36 &    8.36 &   7.52 & 0.49 &  $-$ &  6.83 &6.88 & 7.42 &  $-$ &   $-$ &  $-$ &  $>9\arcsec$ \\
S103 & 18 34 43.92 &$-$07 14 47.0 &   	 9.76  &    8.07 &   7.34 & 0.38 & 6.97 & 6.79 &6.73  & \multicolumn{5}{c}{none within $10\arcsec$}  \\
S104 & 18 35 02.42 &$-$07 09 33.4 &   	 10.26 &    8.40 &   7.58 & 0.38 &7.22 &6.96& 6.89&  6.53&   $-$ &   $-$ &  $-$ &  $1\farcs4$  \\
S105 & 18 35 07.59 &$-$07 18 47.4 &   	 10.33 &    8.36 &   7.59 & 0.58 &7.13 &6.89& 6.95&  5.06 &4.26 &5.30&  $-$ &  $5\farcs9$ \\
S106 & 18 35 06.67 &$-$07 18 55.2 &   	 10.15 &    8.24 &   7.41 & 0.42& 7.09& 6.78& 6.82& \multicolumn{5}{c}{none within $10\arcsec$}\\
S107 & 18 34 27.37 &$-$07 16 19.1 &   	 9.74  &    7.94 &   7.10 & 0.28 & 6.86 & 6.53& 6.53&  5.87& 5.34 &  $-$ &  $-$ &  $3\farcs1$\\
S108 & 18 34 28.66 &$-$07 09 57.4 &      9.94	&    8.02 &   7.17 &0.15 & 6.88  &6.51 & 6.52&  6.54& $-$  &  $-$ &  $-$ &  $1\farcs3$ \\
S109 & 18 34 50.86 &$-$07 18 11.9 & 	 10.45	&    8.61 &   7.76 &0.31& 7.43 &7.13 & 7.11&  6.08&   $-$ &   $-$ &  $-$ &  $5\farcs1$   \\ 
 \hline
\end{tabular}
	\begin{list}{}{}
\item[]$^{a}$  Co-ordinates
and near-IR magnitudes are from 2MASS, with mid-IR ($\sim4$--25~$\mu$m)
magnitudes from the Galactic plane surveys of GLIMPSE/{\it Spitzer}
\citep{benjamin} and the {\it Midcourse Source
  Experiment (MSX)}  \citep{egan}. 
\item[]$^{b}$ The nominal positions of {\it MSX} sources
have $1\farcs5$ uncertainties. Offsets much larger than $3\arcsec$ are
then likely to indicate random superpositions.
\item[]$^{c}$ None of the candidate cluster members is detected by {\it Spitzer} at 3.6$\mu$m. 
\item[]$^{d}$The $J$ magnitude for S8 has quality flag E, indicating a poor fit of the PSF.
\end{list} 
\end{table*}

 \begin{table*}
\label{tab:sgs}
\caption{Summary of the stellar properties of the 9 RSGs for which spectral classification was possible.}
  \begin{tabular}{lccccccccccc}
        \hline \hline
        ID & Spec & $T_{{\rm eff}}(K)^{a}$ & $(J-K)_{0}$& $(J-K)$& $E(J-K)$&$A_{K}^{c}$ & $M_{K}^{c,d}$ \\
             & Type & &        &      &     \\
        \hline
 S1 & K5\,I &$3840\pm135$ &0.75 & 3.25 & 2.50 &1.68 & $-$9.1 \\
 S2 & K4\,I & $3920\pm112$&0.72 & 3.54 & 2.82 &1.89& $-$8.7   \\
 S3 & M3\,I & $3605\pm147$&0.90 & 3.56 & 2.66 &1.78& $-$9.9   \\
 S4 & M6\,I & $3400\pm150$&1.05$^{b}$ & 4.59 & 3.54 &2.37& $-$10.3 \\
 S5 & K5\,I & $3840\pm135$&0.75 & 3.77 & 3.02 &2.02&$-$9.9\\
 S6 & M2\,I & $3660\pm127$&0.87 & 3.83 & 2.96 &1.98&$-$9.7\\
 S7 & K5\,I & $3840\pm135$&0.75 & 3.40 & 2.65 &1.78&$-$8.8\\
 S8 & M2\,I & $3660\pm127$&0.87 & 3.79 & 2.92 &1.94&$-$9.4\\
 S9 & M1\,I & $3745\pm117$&0.85 & 3.34 & 2.49 &1.67&$-$9.3\\
   \hline			   
   \end{tabular}
	\begin{list}{}{}
\item[]$^{a}$ Assumed from the calibration of \citet{levesque},
  following \citet{davies08}.
\item[]$^{b}$ Extrapolated from the calibration.
\item[]$^{c}$ Typical errors in $A_{K}$ and $M_{K}$ are 0.2~mag (see text for discussion).
\item[]$^{d}$ Assuming a distance of 6.6~kpc, identical to RSGC1.
\end{list} 	   
\end{table*}

Interestingly, Table~\ref{tab:props} shows that, amongst the confirmed
RSGs, the three stars with late spectral types are the only ones 
detected in all MSX bands, though their dereddened [A$-$C] colours do
not provide immediate evidence for colour excesses indicative of a
large dust envelope. \citepalias[cf.][]{davies07}. However, the very high $E(J-K_{{\rm S}})$ and $E(H-K_{{\rm S}})$ colours of S4, suggest intrinsic
extinction, indicative of circumstellar material.

\subsection{Reddening and age}

The lack of kinematic data also complicates a determination of the
distance to the cluster. RSGs span a wide range of  luminosities   
($\log(L_{{\rm bol}}/L_{\odot})\sim4.0$--5.8; \citealt{meynet00}), and
therefore absolute magnitudes cannot be inferred from the approximate spectral
types. In addition, the extinction in this direction is very high.
\citet{davies08} derive 
$A_{K_{{\rm S}}}=2.6$ for the nearby RSGC1. We make a quick estimation
of the reddening to Alicante~8 by using the intrinsic colours of RSGs from
\citet{elias85}. We take the values for luminosity class Iab stars,
but, considering the huge values of the reddening and the uncertainty
in the spectral type, this choice is unlikely to be the main
contributor to dispersion. We note that the intrinsic colours of \citet{elias85}
are in the CIT system, but again this effect is unlikely to result in
a major contribution to dispersion. Individual values are listed in
Table~2. 

The main source of errors in the calculation of $A_K$ (and so $M_K$)
is the uncertainty in the spectral type calibration from EW$_{{\rm CO}}$, which \citetalias{davies07} estimate at $\pm2$ subtypes. This value is high enough to allow neglecting the uncertainty in the actual value of EW$_{{\rm CO}}$. Rather than propagating this uncertainty through the calculations, we evaluate the total error by constructing a simple Monte Carlo
simulation. For a given set of supergiants, with intrinsic colors taken
from \citet{elias85}, we draw extinctions in the $K$ band from a normal
distribution $N(\mu,\sigma)=(2,0.5)$, representative of the expected range
for our observations, and use them to calculate their reddened colours and
magnitudes. We assign to each of this mock stars an "observed" spectral
type (i.e., their real spectral type plus or minus the expected 2 subtypes). With the correspondent intrinsic colour, we invert the equations to obtain a value for $A_K$ and, from it, the corresponding $M_{K}$. With this procedure, we estimate that the error
in the spectral types translates into a $\pm0.15$ difference in $A_K$,
using a single colour, and $0.1$ when averaging the extinction derived
from $(H-K_{{\rm S}})$ and $(J-K_{{\rm S}}$). Adding in quadrature the typical errors in the observed colours ($0.05$~mag) and in the intrinsic colour ($0.05$~mag), we reach a final figure of $\pm0.2$~mag for every individual determination of $M_{K}$.

We find averages $E(J-K_{{\rm S}})=2.7\pm0.2$ and
$E(H-K_{{\rm S}})=1.02\pm0.07$, where the errors represent the
dispersions in the individual values. We exclude S4 from this
analysis, as its $(J-K_{{\rm S}})$ is almost one mag higher than those
of all other stars, likely indicative of intrinsic extinction. We also exclude S8, as its 2MASS $J$ magnitude is marked as unreliable, though the values obtained for this star are fully compatible with the others and including it does not change the averages significantly.

The ratio of colour excesses $E(J-K)/E(H-K)=2.7$ is fully compatible with the 
standard extinction law \citep[e.g., 2.8 in][]{inde05}. These values
translate into $A_{K_{{\rm S}}}=1.9\pm0.2$, where the uncertainty
reflects the dispersion in individual values and the slight difference
between the values derived from $E(J-K)$ and  $E(H-K)$. The reddening is
thus lower than towards RSGC1, but higher than towards the other two RSG
clusters in the area.

\begin{figure}
\label{fig:extin}
\resizebox{\columnwidth}{!}{\includegraphics[angle=90,clip]{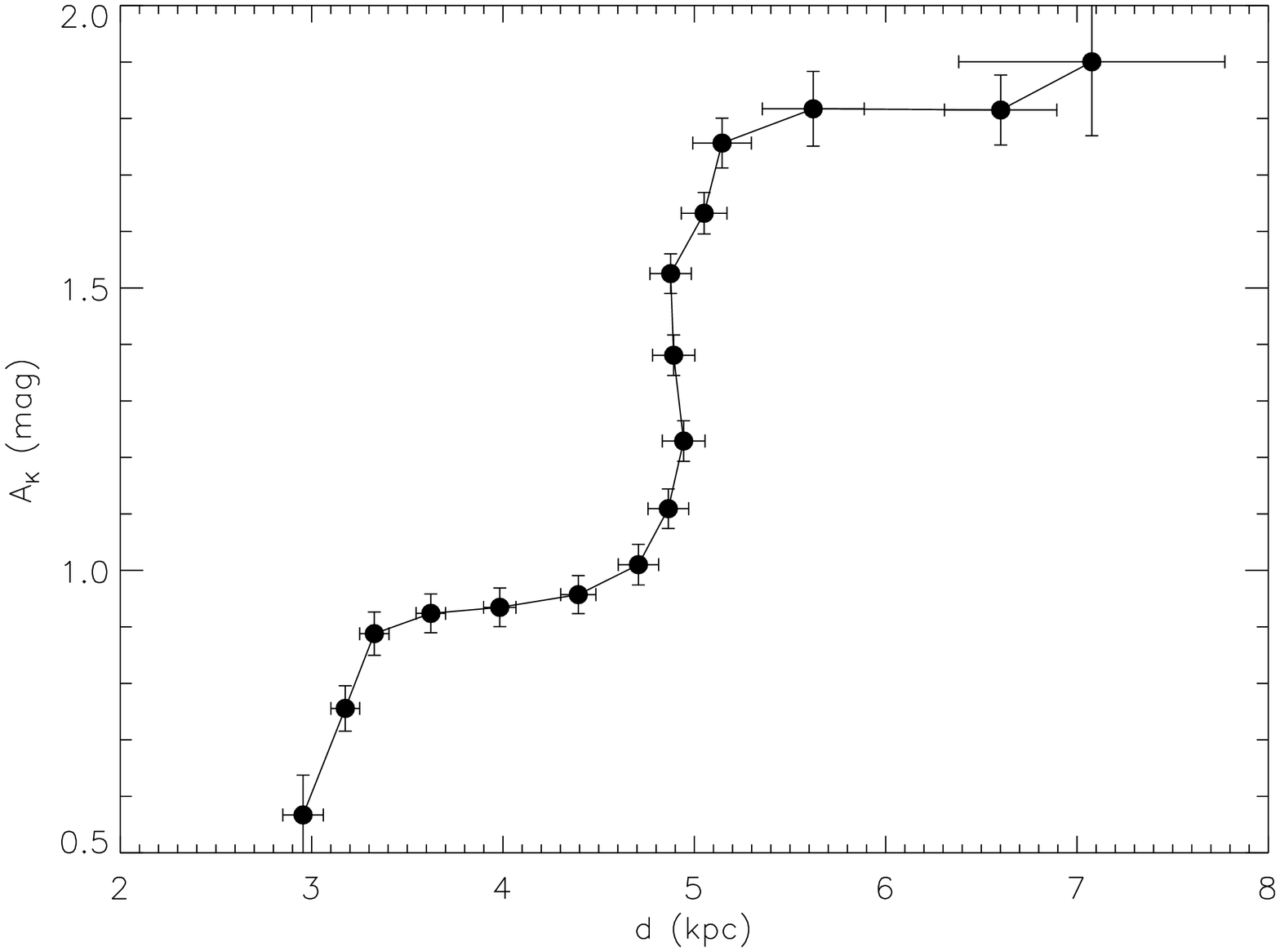}}
\caption{Run of the extinction in the direction to Alicante~8. The data have been obtained by applying the technique of \citet{cabrera05} to the red clump giant population within $20\arcmin$ from S4. The sudden increase in the reddening at $\sim 5$~kpc provides a strong lower limit to the distance to the cluster.}
\end{figure}

We can obtain a firm estimate of the distance to the cluster by studying the distribution of interstellar extinction in this direction. For this, we utilise the  population of red clump giants (with spectral type K2\,III), following the technique of \citet{cabrera05}. This population is seen in the colour-magnitude diagrams as a well defined strip. In the UKIDSS data, we select the giant population within $20\arcmin$ from Star~4, obtaining the results shown in Fig.~6. This radius is chosen in order to keep the $(d,A_K)$ curve representative for the cluster sightline, while providing a number of stars high enough to permit a proper calculation. Decreasing this value to, for example, $10\arcmin$
produces noisier results, but does not change the overall behaviour of
the extinction. As it is clearly seen, most of the extinction along this sightline arises in a small region located at $d\approx 5$~kpc. The values of $A_{K}$ measured for Alicante~8 place the cluster at a higher distance, behind  the extinction wall, $\ga 6$~kpc.

Red supergiants in RSGC1 span $K_{{\rm S}}=5.0-6.2$. Those in
Alicante~8 cover the range $K_{{\rm S}}=6.0-7.1$ (reaching $K_{{\rm S}}=7.4$ if
candidate S13 is confirmed as a member). The range in magnitudes is
approximately the same, but the stars are one magnitude fainter. If we
take into account that the extinction is higher towards RSGC1, we find
that the dereddened magnitudes for stars in Alicante~8  are $\sim1.8$~mag fainter than
those in RSGC1. Given the very high extinction in this region, 
the possibility that Alicante~8 is significantly more distant than
RSGC1 looks very unlikely. Both the distribution of stars in Fig.~3 and the lack of reliable points for $d\ga7$~kpc in Fig.~6 suggest that the reddening reaches very high values at the distance of the cluster. This rise in extinction could be associated to the presence of molecular clouds in the Molecular Ring. We must thus conclude that the RSGs in
Alicante~8 are considerably less luminous than those in RSGC1. Indeed,
if we assume a distance $d=6.6$~kpc, that found for RSGC1
\citep{davies08}, we find absolute $M_{K}$ magnitudes ranging from
$-8.7$ to $-10.3$. This 
range is directly comparable to that seen in RSGC3, and implies an age
of 16--20~Myr for Alicante~8, the age found for RSGC3
\citep{clark09}\fnmsep\footnote{Note that 
  \citet{alexander09} derive a slightly older age (18--24~Myr) for
  RSGC3, based on a fit to isochrones for non-rotating stellar populations by
  \citet{marigo}. The difference is most likely due to the extinction
  laws assumed.}, as opposed to the   
$\sim12$~Myr for RSGC1 \citep{davies08}.

\begin{figure}
\label{fig:theotcmd}
\resizebox{\columnwidth}{!}{\includegraphics[angle=-0,clip]{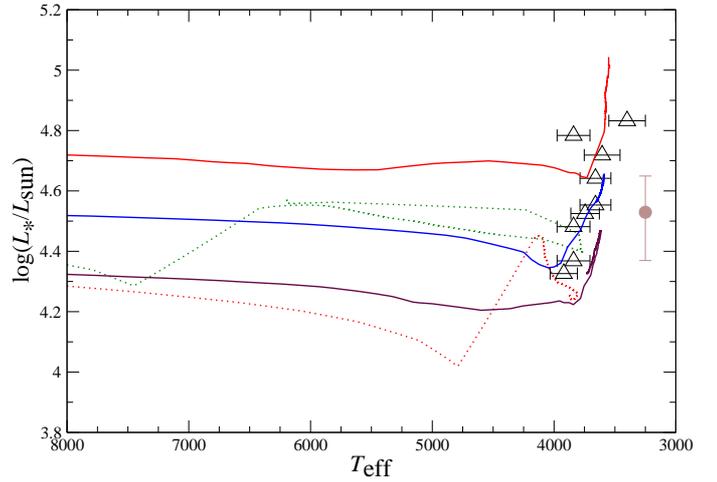}}
\caption{H-R diagram showing the locations of 8 RSGs at the
  cluster centre, with their positions derived from the spectral
  classification, assuming a distance to the
cluster $d=6.6$~kpc. We also plot isochrones from
\citet{meynet00}. The solid lines are the $\log t=7.20$ (16~Myr; 
top, red), $\log t=7.30$ (20~Myr; middle, blue) and $\log t=7.40$ (24~Myr; bottom, black) with high initial
rotation. The dotted lines are the $\log t=7.15$ (14~Myr; top, green)
and  $\log t=7.20$ (bottom, red) isochrones without rotation. Errors
in $\log L_{*}$ due to observational uncertainties and calibration
issues are small when compared to the uncertainty in the cluster
distance ($\Delta D_{{\rm cl}}$);  representative error-bars assuming
an uncertainty of $\pm1\:$kpc are indicated to the right of the figure.}
\end{figure}

To confirm the age derivation, we plot in Fig.~7 the
locations of the RSGs in the theoretical H-R diagram. For this, we
follow the procedure utilised by \citetalias{davies07}. Using the
individual extinctions measured above, we derive absolute $M_{K_{{\rm
      S}}}$ magnitudes for the stars, assuming a distance modulus
$DM=14.1$ ($d=6.6$~kpc). We then use the effective temperature
calibration and bolometric corrections of \citet{levesque} to
derive $T_{{\rm eff}}$ and $L_{*}$ for each object \fnmsep\footnote{We
  note that consistency would perhaps demand that we use the intrinsic
  colours from \citet{levesque}, but we prefer to use the same
  methodology as \citetalias{davies07} in order to ease comparison. Using
  the colours of \citet{levesque} reduces the extinction $A_{K_{{\rm
        S}}}$ by $\sim0.2$~mag, correspondingly decreasing $M_{K_{{\rm
        S}}}$ by slightly more than 0.1~mag, too small a difference
  for any significant impact on the parameters derived.}. In
Fig.~7, we also plot different 
isochrones corresponding to the models of \citet{meynet00}. The
observational temperatures and luminosities of the RSGs are bound by the
$\log t=7.30$ (20~Myr) isochrone for stars without rotation and the
$\log t=7.15$ (14~Myr) isochrone with high initial rotation
($v_{{\rm rot}}=300\:{\rm km}\,{\rm s}^{-1}$). As stars
in the cluster are likely to have started their lives with a range of
rotational velocities, the data are consistent with an age in the
16--20~Myr range. Reducing the distance to the cluster to the nominal
6~kpc adopted by \citet{clark09} for RSGC3 results in a slight
increase of age. In this case, the luminosities of some RSGs are marginally consistent with the high rotation $\log t=7.40$ (24~Myr) isochrone, though the brightest RSGs seem incompatible with this age.

\subsection{The sightline}

The stellar population in the direction to Alicante~8 is very poorly known. The Sagittarius Arm is very sparsely traced by the open clusters NGC~6649 ($\ell=21\fd6$), NGC~6664 ($\ell=24\fd0$) and Trumpler~35 ($\ell=28\fd3$). The reddening to these three clusters is variable, but moderate, with values $E(B-V)\approx1.3$ for Trumpler~35 and NGC~6649 \citep{turner80,majaess08}. The reddening law is compatible with standard ($R=3.0$) over the whole area \citep{turner80}. Around $\ell=28\degr$, \citet{turner80} found several luminous OB supergiants with distances in excess of 3~kpc and reddenings $E(B-V)\approx1.3$, corresponding to $A_{K}=0.4$. This agrees with our determination of  $A_{K}=0.5$ at $d=3$~kpc. The reddening increases steeply between 3 and 3.5~kpc, the expected distance for the Scutum-Crux Arm. It then remains approximately constant until it suffers a sudden and brutal increase around $d=5$~kpc.

As mentioned above a bright star not selected as a candidate member, S101, fell by chance in one of our slits and turns out to be a luminous red star, though not a 
supergiant. Examination of the 2MASS CMDs shows that it is part of a
compact clump of bright stars, which have been marked as green circles in
Fig.~3. These objects, labelled S101--109, are clearly
clumped in both the $(J-K_{{\rm S}})/K_{{\rm S}}$  and  $(H-K_{{\rm S}})/K_{{\rm S}}$
diagrams, at much brighter magnitudes than the field population of red
clump giants, but are uniformly spread over the field studied. We list
their magnitudes in Table~\ref{tab:props}. These stars are too bright to be red clump stars at any distance. Indeed, their average $(J-K)=2.7$ means that, even if they are late M stars, they must be located behind $A_{K}\sim1.0$. In view of this, they could be a population of luminous M giants in the Scutum-Crux Arm, implying typical $M_{K}\approx-6$. They can also be located at the same distance as the extinction wall, but then would have $M_{K}\la-7$, approaching the luminosity of the brightest AGB stars \citep{vanloon05}.

 Cross correlation with the
DENIS catalogue shows that these 
objects are all relatively bright in the $I$ band, while cluster
members are close to the detection limit (with $I\sim17$) or not detected
at all. These again suggests that this population of red giants is closer than the cluster, favouring the Scutum-Crux Arm location. Interestingly, none of these objects has a clear detection in
the {\it MSX} catalogue (Table~\ref{tab:props}), again confirming
their lower intrinsic luminosity.

\subsection{The cluster against the background}

Unfortunately, we cannot find the sequence of unevolved members for Alicante~8 in either 2MASS or UKIDDS photometry. In this respect, it is worth considering the properties of the open cluster NGC~7419, which contains five RSGs and, though moderately extinguished, is visible in the optical. The 2MASS colour-magnitude diagram for NGC~7419 does not show a well defined sequence, in spite of the fact that the field contamination is very small at the magnitudes of the brightest blue members \citep{joshi08}. This is due to differential reddening and the presence of a significant fraction of Be stars amongst the brightest members, which show important colour excesses. With the much higher extinction and field contamination of Alicante~8, its unevolved sequence would be most likely undetectable.

However, we carry out some further tests in order to verify our conclusions. First, we try to estimate the likelihood that the overdensity associated with the cluster may be the result of a random fluctuation. This is very difficult to evaluate, given the very red colours of the stars. The $r=3\arcmin$ circle centred on S4 shows a clear overdensity of bright stars with respect to the surrounding field. The significance of this overdensity depends very strongly on the set of parameters we use to define the comparison population: we could choose just ``bright'' stars (i.e., $K<7$) or add some extra criteria, such as very red colours (e.g., $(J-K)>2$) or $Q$ incompatible with a red clump giant ($Q<0.4$). Depending on the criteria selected, the  $r=3\arcmin$ circle presents an overdensity by a factor 2--3 with respect to the surrounding ($<1\degr$) field. The existence of the cluster, however, is defined by the presence of a very well defined clump of bright stars in the $Q/K_{{\rm S}}$,  $(H-K_{{\rm S}})/K_{{\rm S}}$, and $(J-K_{{\rm S}})/K_{{\rm S}}$ diagrams, which no other nearby $r=3\arcmin$ circle seems to present.

The possibility that Alicante~8 represents a random overdensity of RSGs seems extremely unlikely in view of the rarity of these objects. In order to consider this option, we would have to assume that most stars with $K_{{\rm S}}<7.5$ in the surrounding field are RSGs, leading to a population of hundreds of RSGs for each square degree. The only other possibility of a random fluctuation would be the random coincidence of a small cluster of RSGs with a number of luminous M giants that happen to have the same colours. This also seems very unlikely. The mid-IR colours of all our candidate supergiants suggest that they are not surrounded by dust. If any ot them were M giants without dusty envelopes, their colours would be only a few tenths of a magnitude redder than those of K supergiants, meaning that they would still be highly reddened and should be placed behind the reddening wall at $d\approx6$~kpc. Though some AGB stars can reach very bright magnitudes \citep[$A_{K}\la-8$;][]{vanloon05}, these are very rare objects \citep[e.g.,][for the Magellanic Clouds]{groene09}, descended only from the most massive intermediate-mass stars \citep{mg08}. Therefore such a chance coincidence looks equally unlikely.
 
\section{Discussion}
\label{sec:discu}

The data available reveal that Alicante~8 is a new highly reddened
open cluster in the same area where three others had already been
located. This discovery represents further evidence for the existence
of intense star formation in the region between Galactic longitude
$\ell=24\degr-28\degr$. Sightlines in this direction are believed to
cross the Sagittarius Arm, cross through the Scutum Arm and then hit
the Long Bar close to its intersection with the base of the Scutum
Arm at $\ell\sim27\degr$, at an estimated distance of $\sim6.5$~kpc.
 
This coincidence strongly suggests that the tip of the Bar is
dynamically exciting star formation giving rise to a starburst
region \citep[see discussion in][]{davies07,garzon}. If we take into account
the spatial span covered by the four 
clusters known, 
this would be by far the largest star-forming region known in the
Milky Way. 

An alternative view, based on the distribution of molecular clouds in
radio maps, is that a giant Molecular Ring is located at the end of
the Bar, at a distance $\approx4.5$~kpc from the Galactic Centre. In this
view, our sightline would be cutting through the Ring. We would then
be looking through 
the cross section of a giant star-forming ring, coincident with the
Molecular Ring seen in the radio. between distances $\sim5$
and $\sim8$~kpc from the Sun. In this case, the clusters could be spread in
depth over a distance $\sim3$~kpc, and not necessarily be associated.
As the unevolved population of Alicante~8 cannot be detected, an
estimation of its distance will have to wait for data that can provide
dynamical information. Meanwhile, we will stick to the assumed 6.6~kpc.

Likewise, a direct
estimate of the cluster mass cannot be made. Recent simulations of
stellar populations with a Kroupa IMF \citep{simonw51} indicate that a
population of $10\,000\:M_{\sun}$ at 16--20~Myr should contain 2--5
RSGs. Cruder estimates using a Salpeter IMF, like those in
\citep{clark09}, suggest 8 RSGs for each
$10\,000\:M_{\sun}$. Therefore, based on the membership of at least 8
RSGs, we can estimate that Alicante~8 contains a minimum of $10\,000\:M_{\sun}$
and, if some of the candidates outside the core are confirmed, could approach
$20\,000\:M_{\sun}$. Thus, it seems that it is between half and one
third the mass of RSGC2 and RSGC3, which have similar ages, and may be one of
the ten most massive young clusters known in the Galaxy.

It is thus quite significant that Alicante~8 does not stand out at
all in GLIMPSE mid-IR images, and is only moderately conspicuous over
the crowded field in near-IR images. As a matter of fact, the cluster
would not appear evident to the eye were it not for the presence of a
few foreground objects which, fortuitously, make the clumping of bright
stars more apparent (Fig.~\ref{fig:colour}). 

In the presence of such a rich foreground (and likely background)
population, the detection of massive clusters, even if they are
moderately rich in red supergiants, may be a question of chance
coincidence with a void in the distribution of bright foreground stars
or a hole in the extinction. In this respect, it is worth noting that
RSGC1 stands out because of its youth (and hence the intrinsic
brightness of its RSGs), while Stephenson~2, apart from being
extraordinarily rich in RSGs, is located in an area of comparatively
low extinction. 

Alicante~8 is located $\approx16\arcmin$ away from RSGC1. If the two
clusters are located at a common distance of 6.6~kpc, this angular separation
represents a distance of 31~pc, consistent with the size of cluster
complexes seen in other galaxies \citep{nate05}. Even if Stephenson~2
(which would be located at $\sim 100$~pc from RSGC1 in the opposite
direction to Alicante~8) is also physically connected, the distances
involved are not excessive. The inclusion of RSGC3, located at 400~pc,
in the same starburst region is more problematic, requiring it to be a
giant star-formation region. At such distance, the possibility of triggered star formation (in any direction) seems unlikely, but large complexes may form caused by external triggers, as is likely the case of W51 (\citealt{simonw51}; Parsons et al., in preparation).

\citet{lopez} have reported the existence of a diffuse population of
RSGs in this area, while \citetalias{davies07} detect several RSGs around
Stephenson~2 with radial velocities apparently incompatible with
cluster membership. Therefore the actual size of the star forming
region still has to be determined. The age difference between
Alicante~8 and RSGC1 is small, but the Quartet cluster, with an age
between 3 and 8~Myr is also located in the same area (about
$20\arcmin$ due East from Alicante~8), at about the same
distance \citep{messi09}. Relatively wide age ranges ($\sim 5$~Myr)
are common in cluster complexes. Examples are the central cluster in
30~Dor and its periphery \citep{walborn02} or the several regions in
W51 \citep{simonw51}. 

We have searched for other objects of interest in the immediate
vicinity of Alicante~8, but no water masers or X-ray sources are known
within $10\arcmin$ of the cluster. The lack of young X-ray binaries,
though not remarkable over such a small area, becomes intriguing when
the whole area containing the RSG clusters is considered
\citep[cf.][]{clark09}.

\section{Conclusions}

Alicante~8 contains at least 8 RSGs. If a distance of 6.6~kpc, common
to the other RSGCs, is assumed, its age is $16-20$~Myr. The presence of
these 8 RSGs would then imply a mass in excess of
$10\,000\:M_{\sun}$, which could approach $20\,000\:M_{\sun}$ if the
candidate members are confirmed.

The discovery of a fourth cluster of red supergiants in a small
patch of the sky confirms the existence of a region of enhanced star
formation, which we will call the Scutum Complex. As the properties of
the four known clusters do not rule out the presence of many other
smaller clusters, we are faced with the issue of determining the true
nature and extent of this complex. Assuming a common distance for all
clusters results in a 
coherent picture, as they are all compatible with a narrow range of
ages (between $\sim12$ and $\sim20$~Myr), showing a dispersion typical
of star-forming complexes. However, the spatial extent of this complex
should be several hundred parsecs, rising questions about how such a
massive structure may have arisen in our Galaxy.

Further spectroscopic studies, combined with precise radial velocity
measurements, will be necessary to confirm the membership of candidate
RSGs in the field of Alicante~8 and provide a better estimate of its
mass. Radial velocities  and accurate parallaxes will
also be necessary to establish the actual spatial and temporal
extent of this putatively giant starburst region in our own Galaxy.

\begin{acknowledgements}

 We thank the referee, Dr. Ben Davies, for his useful suggestions, which led to substantial improvement. We thank Antonio Flor\'{\i}a for the enhancement effects in the three-colour image of the cluster. 

 The WHT is operated on the island of La
Palma by the Isaac Newton Group in the Spanish Observatorio del Roque
de los Muchachos of the Instituto de Astrof\'{\i}sica de Canarias. 
We thank the ING service programme for their invaluable
collaboration. In particular, we thank M.~Santander for his support in
preparing the observations.

This research is partially supported by the Spanish Ministerio de
Ciencia e Innovaci\'on (MICINN) under
grants AYA2008-06166-C03-03 and CSD2006-70, and by the Generalitat Valenciana under grant ACOMP/2009/164. JSC acknowledges support
from an RCUK fellowship. SMN is a researcher of the Programme Juan de
la Cierva, funded by the MICINN. 

 UKIDSS uses the
UKIRT Wide Field Camera (WFCAM; Casali et al. 2007) and a photometric
system described in \citet{hewett06}. The pipeline processing and
science archive are described in \citet{hambly08}.

 This publication
makes use of data products from 
the Two Micron All 
Sky Survey, which is a joint project of the University of
Massachusetts and the Infrared Processing and Analysis
Center/California Institute of Technology, funded by the National
Aeronautics and Space Administration and the National Science
Foundation.

\end{acknowledgements}

{}

\begin{thebibliography}{}

\bibitem[Alexander et al.(2009)]{alexander09} Alexander, M.J.,
  Kobulnicky, H.A., Clemens, D.P., et al. 2009, AJ, 137, 4824

\bibitem[Bastian et al.(2005)]{nate05}
Bastian, N., Gieles, M., Efremov, Yu.N., \& Lamers, H.J.G.L.M. 2005, A\&A, 443, 79


\bibitem[Benjamin et al.(2003)]{benjamin}
Benjamin, R.A., Churchwell, E., Babler, B.L., et al. 2003, PASP, 115, 953

 \bibitem[Cabrera-Lavers et al.(2005)]{cabrera05} Cabrera-Lavers, A., Garz\'on, F., \& Hammersley, P.L. 2005, A\&A, 433, 173

 \bibitem[Cabrera-Lavers et al.(2008)]{cabrera08} Cabrera-Lavers, A., Gonz\'alez-Fern\'andez, C., Garz\'on, F., et al. 2008, A\&A, 491, 781   

 \bibitem[Casali et al.(2007)]{casali07} Casali, M., Adamson, A.,
  Alves de Oliveira, C., et al. 2007, A\&A, 467, 777 


\bibitem[Clark et al.(2005)]{clark05}
Clark, J. S., Negueruela, I., Crowther, P. A., \& Goodwin, S. P. 2005, A\&A, 
434, 949 

\bibitem[Clark et al.(2009a)]{clark09}
Clark, J. S., Negueruela, I., Davies, B., et al. 2009a, A\&A, 
498, 109

\bibitem[Clark et al.(2009b)]{simonw51}
Clark, J. S., Davies, B., Najarro, F., et al. 2009b, A\&A, 
504, 429


\bibitem[Cotera et al.(1996)]{cotera}
Cotera, A.S., Erickson, E.F., Colgan, S.W.J., et al., 1996, ApJ, 461, 
750

\bibitem[Davies et al.(2007)]{davies07}
Davies, B., Figer, D.F., Kudritzki, R.-P., et al. 2007, ApJ, 671, 781 (D07)

\bibitem[Davies et al.(2008)]{davies08}
Davies, B., Figer, D.F., Law, C.J. et al. 2008, ApJ, 676, 1016 



\bibitem[Egan et al.(2001)]{egan}
Egan, M. P., Price, S. D., Gugliotti, G. M., 2001, BAAS, 34, 561

\bibitem[Elias et al.(1985)]{elias85}
Elias, J.H., Frogel, J.A., \& Humphreys, R.M. 1985, ApJ, 57, 91
 
\bibitem[Figer et al.(1999)]{figer99}
Figer, D.F., McLean, I.S., \& Morris, M. 1999, ApJ, 514, 202



\bibitem[Figer et al.(2006)]{figer06}
Figer, D.F., MacKenty, J.W., Robberto, M., et al. 2006, ApJ, 643, 1166 

\bibitem[Garz\'on et al.(1997)]{garzon}
Garz\'on, F., L\'opez-Corredoira, M., Hammersley, P., et al. 1997, ApJ, 491, L31


\bibitem[Gonz\'alez-Fern\'andez et al.(2008)]{carlos08} Gonz\'alez-Fern\'andez,
  C., Cabrera-Lavers, A., Hammersley, P.L., \& Garz\'on, F. 2008, A\&A, 479, 131





\bibitem[Groenewegen et al.(2009)]{groene09} Groenewegen, M.A.T., Sloan, G.C., Soszy\'{n}ski, I., \& Petersen, E.A. 2009, A\&A, 506, 1277 

\bibitem[Hambly et al.(2008)]{hambly08} Hambly, N.C., Collins, R.S.,
  Cross, N.J.G., et al. 2008, MNRAS, 384, 637

\bibitem[Hewett et al.(2006)]{hewett06} Hewett, P.C., Warren, S.J.,
  Leggett, S.K., \& Hogkin, S.T. 2006, MNRAS, 367, 454

\bibitem[Humphreys \& McElroy(1984)]{hme84} Humphreys, R.M., \& McElroy, D.B. 1984, ApJ, 284, 565 

\bibitem[Joshi et al.(2008)]{joshi08} Joshi, H., Kumar, B., Singh, K.P., et al. 2008, MNRAS, 391, 1279

\bibitem[Indebetouw et al.(2005)]{inde05} Indebetouw, R., Mathis,
  J.S., Babler, B.L., et al. 2005, ApJ, 619, 931

\bibitem[Krabbe et al.(1995)]{krabbe}
Krabbe, A., Genzel, R., Eckart, A., et al., 1995, ApJ, 447, L95

 \bibitem[Lawrence et al.(2007)]{lawrence07} Lawrence, A., Warren, S.J.,
  Almaini, O., et al. 2007, MNRAS, 379, 1599  

\bibitem[Levesque et al.(2005)]{levesque} Levesque, E.M., Massey, P.,
  Olsen, K.A.G., et al. 2005, ApJ, 628, 973 

\bibitem[van Loon et al.(2005)]{vanloon05} van Loon, J.Th., Cioni, M.-R. L., Zijlstra, A.A., \& Loup, C. 2005, A\&A, 438, 273

\bibitem[L\'opez-Corredoira et al.(1999)]{lopez}
L\'opez-Corredoira, M., Garz\'on, ~F., Beckman, J.E., et al. 1999, AJ, 118, 381

\bibitem[Lucas et al.(2008)]{lucas08} Lucas, P.W., Hoare, M.G.,
  Longmore, A., et al. 2008, MNRAS, 391, 136

\bibitem[Majaess et al.(2008)]{majaess08} Majaess, D.J., Turner, D.G., \& Lane, D.J. 2008, MNRAS 390, 1539

\bibitem[Marigo \& Giradi(2008)]{mg08}Marigo, P., \& Girardi, L. 2008, A\&A, 469, 239

\bibitem[Marigo et al.(2008)]{marigo}Marigo, P., Girardi, L.,
  Bressan, A., et al. 2008, A\&A, 482, 883


 \bibitem[Messineo et al.(2009)]{messi09}
Messineo, M., Davies, B., Ivanov, V.D., et al. 2009, ApJ, 697, 701

\bibitem[Meynet \& Maeder(2000)]{meynet00}
Meynet, G., \& Maeder, A. 2000, A\&A, 361, 101

\bibitem[Nagata et al.(1995)]{nagata}
Nagata, T., Woodward, C.E., Shure, M., \& Kobayashi, N., 1995, AJ, 109, 1676

\bibitem[Negueruela \& Schurch(2007)]{ns07} Negueruela, I., \& Schurch,
  M.P.E. 2007, A\&A, 461, 431

\bibitem[Rathborne et al.(2009)]{rathborne09}
Rathborne, J.M., Johnson, A.M., Jackson, J. M., et al. 2009, ApJS, 182, 131

\bibitem[Rayner et al.(2009)]{rayner09} Rayner, J.T., Cushing, M.C., \& Vacca, W.D. 2009, ApJS, 185, 289

\bibitem[Strai\v{z}ys \& Lazauskait\.{e}(2009)]{straizys09}
Strai\v{z}ys, V., \& Lazauskait\.{e}, R. 2009, Balt. Astr., 18, 19


\bibitem[Smartt et al.(2009)]{smartt}
Smartt, S.J., Eldridge, J.J., Crockett, R.M., \& Maund, J.R.
2009, MNRAS, 395, 1409

\bibitem[Turner(1980)]{turner80}
Turner, D.G., 1980, ApJ, 240, 137



\bibitem[Vacca et al.(2003)]{vacca03}
Vacca, W.D., Cushing, M.C., \& Rayner, J.T.
2003, PASP, 115, 389

\bibitem[Walborn et al.(2002)]{walborn02}
Walborn, N.R., Ma\'{\i}z-Apell\'aniz, J., \& Barb\'a, R.H., 2002,
AJ, 124, 1601 


\end{thebibliography}
\end{document}